\documentclass[twocolumn,showpacs,preprintnumbers,amsmath,amssymb]{revtex4}
\usepackage{graphicx}

\begin{document}

\title{Screening and inplane magnetoresistance of anisotropic two-dimensional gas}

\author{V.\,S.\,Khrapai\/\thanks{e-mail: dick@issp.ac.ru}}

\affiliation{Institute of Solid State Physics, Chernogolovka,
Moscow District 142432, Russia}

\date{\today}

\begin{abstract}

In order to split the influence of the orbital and spin effects on the inplane magnetoresistance
of a quasi two-dimensional (2D) gas we derive its linear response function and dielectric function
for the case of anisotropic effective mass. This result is used for the calculation of elastic transport
relaxation time of a quasi two dimensional system in a
parallel magnetic field. The relaxation time is proved
to be isotropic in the low density limit  for the case of charged
impurity scattering, allowing to separate the two contributions.
\end{abstract}

 \maketitle

Among a variety of experimentally used two dimensional
semiconductor structures some possess the anisotropy of Fermi
surface originating from that of a bulk material, including n-type
Si-MOSFET's on other than (100) surfaces~\cite{afs}, AlAs
heterostructures~\cite{shayeganAl}, p-type Si and GaAs
structures~\cite{kittel}. The transport properties of such
anisotropic semiconductors has been addressed
largely~\cite{herring,samoilovich,tokura} and it is well known, that the Fermi
surface anisotropy gives rise in
 general to the anisotropy of relaxation time, even if the scattering
 potential is isotropic~\cite{samoilovich}.

Another possibility is the externally introduced anisotropy
through the application of parallel to a quasi 2D layer magnetic
field, known to deform a Fermi surface~\cite{merkt} due to the so
called orbital effect~\cite{sarma}. In this case, however, there
is additionally a coupling of parallel field to the particles'
spins leading to the partial spin polarization of the
system~\cite{Yoon,shayegan}
 and subsequent change of screening~\cite{dolgold} (so called spin effect).
 Although substantial interest~\cite{shayeganAl,Yoon,khrapai,shayegan,liang,zhu,tutuc2003} has been
 recently paid to the longitudinal magnetoresistance (MR) studies of quasi two
 dimensional systems, it remained unclear so far how the two effects play
 together~\cite{khrapai}.

In this letter we address the screening properties of the anisotropic 2D gas
deriving the linear response function for the
case of elliptic Fermi surface. This result is then used to evaluate the transport
relaxation time $\tau$ for elastic charged impurities scattering, which,
surprisingly, turns out to be isotropic in the low density limit.
Furthermore, for partially spin polarized anisotropic 2D system the relaxation time for each spin subband
is also shown to be isotropic, allowing finally to reach the main result of the paper --- to separate
the influence of the orbital and spin effects on the longitudinal
magnetoresistance of a diluted quasi 2D gas.

In the following we utilize the simplest possible representation
of anisotropy --- the elliptic Fermi surface in the effective mass
approximation. The interactions are treated in the random phase
approximation (RPA), so that the screened linear response function
equals to linear response of a free particles gas~\cite{pines}:
\begin{equation}
\chi^{scr.}(\mathbf q,0)=g_vg_s\sum_{\mathbf
k}\left|\left(\rho_q\right)_{n0}\right|^2\frac{f_0(E_{\mathbf
k})-f_0(E_{\mathbf k+\mathbf q})}{E_{\mathbf k}-E_{\mathbf
k+\mathbf q}-i0}\,, \label{chi}
\end{equation}
where $E_{\mathbf k}$ is kinetic energy of a quasiparticle with
momentum $\mathbf k$, $f_0(E_\mathbf k)$ -- the zero temperature
Fermi-Dirac distribution function, $g_v,g_s$ --- the valley and
spin degeneracies. The excited state $|n\rangle$ contains a single
pair of a quasiparticle with momentum $\mathbf k+\mathbf q$ and a
quasihole with momentum $\mathbf k$, and the matrix element of a
density fluctuation operator $\left(\rho_q\right)_{n0}$ is equal
to unity. The last property originates entirely from the Bloch
type of Hamiltonian eigenfunctions in the effective mass
approximation, similar to the isotropic case~\cite{pines}.

We perform the following change of coordinates to rewrite the
integral~(\ref{chi}) in a spherically symmetric form:
\begin{equation}
\begin{array}{cc}
 k_x \rightarrow k_x^F\tilde k_x \quad& \quad q_x \rightarrow k_x^F\tilde q_x \\
 k_y \rightarrow k_y^F\tilde k_y\quad &\quad q_y \rightarrow k_y^F\tilde q_y
\end{array} \label{coords}
\end{equation}
The kinetic energy depends solely upon the length of the distorted wave vector $\tilde{\mathbf  k}$ --- $E_{\mathbf
k}=E_F((k_x/k_x^F)^2+(k_y/k_y^F)^2)=E_F|\tilde{\mathbf k}|^2$ and
for integral~(\ref{chi}) we have:
\begin{equation}
 \chi^{scr.}(\mathbf
 q,0)=g_vg_s\frac{k_x^Fk_y^F}{4\pi^2E_F}\int\frac{f_0(E_{\tilde{\mathbf k}})-f_0(E_{{\tilde{\mathbf k}}
+\tilde{\mathbf q}})}{-2{\tilde{\mathbf k}\cdot\tilde{\mathbf q}
}-\tilde{\mathbf q} ^2-i0}d^2{\tilde{\mathbf k}} \label{chi1}
\end{equation}
In view of spherical symmetry the integral value in~(\ref{chi1})
is invariant to rotation of vector $\tilde{\mathbf q} $, hence the
linear response function depends upon its length $\tilde q$ only.
The integral in~(\ref{chi1}) would be the same for isotropic Fermi
surface as well, thus the only difference from Stern's linear
response function~\cite{stern} is due to the normalizing prefactor
proportional to the density of states of anisotropic Fermi gas
$D=g_vg_s\sqrt{m_xm_y}/2\pi\hbar^2$. Finally, we get for $\chi$:

\begin{equation}
\begin{array}{c}
 \chi^{scr.}(\mathbf q,0)=-D\cdot\phi(\tilde q)\,, \\
\\
\phi(\tilde q)=\left\{\begin{array}{l}
  1, \quad \tilde q<2 \\
  1-(1-4/\tilde q^2)^{1/2},  \quad \tilde q\ge2
\end{array} \right. \end{array}
\label{chi2}
\end{equation}
The linear response function~(\ref{chi2}) of anisotropic system
depends upon the direction of perturbation wave vector $\mathbf q$
through the variable $\tilde q
=((q_x/k_x^F)^2+(q_y/k_y^F)^2)^{1/2}$, so that the screening does
become anisotropic, in contrast to the case of isotropic Fermi
surface~\cite{stern}. Note, however, that this anisotropy is the
same as that of kinetic energy as a function of momentum, since
$\tilde q^2=E_{\mathbf q}/E_F$.

At the same time we are able to find the RPA static dielectric
function~\cite{pines}:
\begin{equation}
 \varepsilon(\mathbf
q,0)=1-V(q)\cdot\chi^{scr.}(\mathbf
q,0)=1+\frac{\,q_{TF}}{q}\,\phi(\tilde q)\label{epsilon}
\end{equation}
where $V(q)=2\pi e^2/q$ is the 2D Fourier transform of the bare
Coulomb interaction potential, $q_{TF}=2\pi e^2D$ --- the
Thomas-Fermi screening parameter~\cite{afs}.

We now turn to the application of our results to the transport
properties of anisotropic 2D Fermi gases.
Expression~(\ref{epsilon}) for dielectric function enables one to
find the elastic transport relaxation time. In general
$\tau(\mathbf k )$ is anisotropic and calculation of this could be
a rather complicated procedure~\cite{tokura,ziman}. As we show below, for
the case of screened charged impurity scattering, $\tau$ is still isotropic
in the low density regime and can be obtained analytically.
One has for elastic scattering transport
relaxation time:
\begin{equation}
\frac{1}{\tau(\mathbf
k)}\propto\int\left(\frac{V(q)}{\varepsilon(\mathbf
q,0)}\right)^2\left(1-\frac{\mathbf v_{\mathbf k }\cdot\mathbf
v_{\mathbf k^\prime}}{\mathrm v_{\mathbf k
}^2}\right)\delta(E_{\mathbf k^\prime}-E_F)\,d^2\mathbf
k^\prime\,, \label{time1}
\end{equation}
where $\mathbf v_{\mathbf k}=\hbar^{-1}dE_\mathbf k/d\mathbf k$ is
the particle's group velocity. Note that the relaxation time
isotropy is already implicit in this expression and is verified in what
follows. First term in the integrand is the square of a scattering
matrix element in the Born approximation~\cite{samoilovich}, where we have neglected all the form-factors of
the real Coulomb interaction between a quasi 2D electron (hole)
and a charged impurity~\cite{afs}. Applying the change of
coordinates~(\ref{coords}) to the $\mathbf k^\prime$-space we find
from~(\ref{epsilon}): $V(q)/\varepsilon(\mathbf q,0)=2\pi
e^2(q+q_{TF}\phi(\tilde q))^{-1}$ which in the low density limit
($k_x^F,k_y^F\ll q_{TF}$) reduces to~\cite{remark2}:
\begin{equation}
V(q)/\varepsilon(\mathbf q,0)=2\pi e^2q_{TF}^{-1}\phi^{-1}(\tilde q)
\label{lowdens}
\end{equation}

The second term in~(\ref{time1}) accounts for a loss of initial
velocity in a scattering event and is similar to the factor
$1-\cos(\widehat{\,\mathbf k,\mathbf k^\prime})$ in the isotropic
case~\cite{ziman}. Changing the coordinates to polar ones $\tilde
k_x,\tilde k_y\rightarrow\tilde k,\theta$ we rewrite this in terms
of a scattering angle $\theta$: $1-\mathbf v_{\mathbf k
}\cdot\mathbf v_{\mathbf k^\prime}/\mathrm v_{\mathbf k
}^2=1-\cos\theta+ A(\mathbf k)\sin\theta\,,$ where number
$A(\mathbf k)$ depends upon the initial particle momentum $\mathbf
k$. Hence in the low density limit we find:
\begin{equation}
\frac{1}{\tau(\mathbf
k)}\propto\frac{1}{g_vg_sD}\int_{-\pi}^{\pi}\phi^{-2}(\cos\theta)(1-\cos\theta)\,d\theta\,,
\label{time2}
\end{equation} since $\tilde q=(2-2\cos\theta)^{1/2}$, and the odd
in $\theta$ part of the integrand gives no contribution to the
integral. This final expression for the elastic scattering time in
the low density limit is essentially the same as in the isotropic
case, the only difference represented by the reciprocal density of
states $D^{-1}$ in the prefactor. This finding has two important
consequences: in the low density limit, within the elliptic
deformation of the Fermi surface, the transport relaxation time
(i) remains isotropic and (ii) increases proportionally to the
density of states~$D$~\cite{remark}.

Comparing to previous studies we find that our result for the
relaxation time isotropy recovers the one derived earlier for the
short range scatterers~\cite{tokura}, since for the case of zero
spin polarization considered so far the relevant Fourier
components of the screened impurity potential~(\ref{lowdens}) do
not depend on wave vector at all, according to~(\ref{chi2}). The
predicted increase of the relaxation time with the Fermi surface
deformation implies of course the increase of conductivity, i.e. the negative magnetoresistance in the
parallel to field direction caused by the orbital effect, as the
 effective mass in this direction remains unchanged~\cite{merkt}. This is in contrast to positive
MR found for orbital effect in Ref.~\cite{sarma}, where the change
of screening has been neglected. Our prediction could be easily verified
in the experiments on wide quantum wells, where the inplane magnetoresistance
is mostly due to the orbital effect.

As was mentioned earlier, apart from the Fermi surface deformation
the parallel magnetic field couples to particles' spins, resulting
in a partial spin polarization of a system. Similar to the
isotropic case~\cite{dolgold} there is no more  a single Fermi
surface, but different ones for different spin projections onto
magnetic field axis. We would like to treat only the spin
conserving processes, which means that the linear response of such
partially polarized system is simply a sum of the responses from
different spin subbands~\cite{dolgold}. Similar to the case of unpolarized system
(expression~(\ref{chi2})), the anisotropy of the linear response of partially polarized
system is again the same as that of the kinetic energy as a function of wave vector
$\mathbf q$, which leads to the isotropy of the relaxation time
for each spin subband in the limit of low density, as we show
below.

Calculating the transport relaxation times $\tau^{\uparrow,
\downarrow}$ one should write the integrals of type~(\ref{time1})
for major ($\uparrow$) and minor ($\downarrow$) spin subbands
separately. In the low density limit, when the Fermi wave vectors
of both subbands satisfy
$k_x^{F\uparrow,\downarrow},k_y^{F\uparrow,\downarrow}\ll q_{TF}$,
the Fourier image of the screened impurity potential $V(\mathbf
q)/\varepsilon(\mathbf q)$ has the anisotropy of the linear response
function and kinetic energy, according to~(\ref{lowdens}). This
means that the change of coordinates~(\ref{coords}) applied to the
integrands leads to the same expressions for $\tau^{\uparrow,
\downarrow}$ as one gets in the isotropic case~\cite{dolgold}.
Thus the relaxation times are isotropic, the only effect of
anisotropy been again to normalize the absolute value of
$\tau^{\uparrow, \downarrow}$ through the density of states
dependent prefactor~(\ref{time2}). We are now able to write down
the conductivity tensor of the partially polarized anisotropic
system:
\begin{equation}
 \hat\sigma=\hat\sigma^\uparrow+\hat\sigma^\downarrow=n_Se^2\tau_0\hat
 m^{-1}\frac{D}{D_0}F_{DG}(\xi), \label{sigma}
\end{equation}
where $\hat\sigma^\uparrow,\hat\sigma^\downarrow$ are the major
and minor spin subbands conductivity tensors, $e,n_S,\tau_0,D_0$
--- respectively, electron charge, density of 2D particles and
the zero field isotropic relaxation time and the density of states.
$D$ and $\hat m$ are respectively the density of states and the
effective mass tensor in magnetic field. The last term
in~(\ref{sigma}) stands for the Dolgopolov-Gold's calculated
change of the conductivity of isotropic system as a function of
its degree of spin polarization
$\xi=(n_\uparrow-n_\downarrow)/(n_\uparrow+n_\downarrow)$~\cite{dolgold}.
Note that for the case of a quasi 2D system with zero field anisotopic mass this result is valid only
for a parallel field applied along the main axes of symmetry, otherwise the Fermi surface loses its ellipticity in magnetic field.
Let us show finally how this simple expression allows to separate
immediately the contributions from spin and orbital effects on the
longitudinal magnetoresistance of a diluted quasi 2D system.

We focus on the recent magnetoresistance studies of the 2D electron gas of AlGaAs/GaAs heterostructure~\cite{khrapai}.
This system is isotropic in zero magnetic field, hence,
according to~(\ref{sigma}) the anisotropy of experimental
MR~\cite{khrapai} is due to the effective mass change in a
perpendicular to magnetic field direction~\cite{sarma}:
$$m_\perp/m_0=\rho_\perp(B)/\rho_\parallel(B)\,,$$ where $\perp$
and $\parallel$ mark the resistances measured in perpendicular and
parallel to inplane magnetic field directions, respectively. The
spin effect contribution in~(\ref{sigma}) is given thus by:
\begin{equation}
F^{-1}_{DG}(\xi)=\sqrt{\frac{\rho_\perp(B)\cdot\rho_\parallel(B)}{\rho_\perp(B=0)\cdot\rho_\parallel(B=0)}}
\label{spin}
\end{equation}
The degree of spin polarization depends upon both the Zeeman
energy $g\mu_BB$ and the effective mass at a given field value:
$\xi(B)=(E_{F\uparrow}-E_{F\downarrow})/(E_{F\uparrow}+E_{F\downarrow})=
g\mu_BB\cdot D/n_S$, where $E_{F\uparrow,\downarrow}$ are the
kinetic parts of the Fermi energy for two spin subbands, $g,\mu_B$ --- the Land\'e factor and Bohr magneton. Equivalently the last
equation reads: $\xi(B)={B/B^0_P\cdot\sqrt{m_\perp(B)/m_0}}$,
where $B^0_P=2E^0_F/g\mu_B$ is the full spin polarization field in
the absence of orbital effect~\cite{dolgold}.

\begin{figure}[hb]
\begin{center}
\includegraphics[width=8cm,clip]{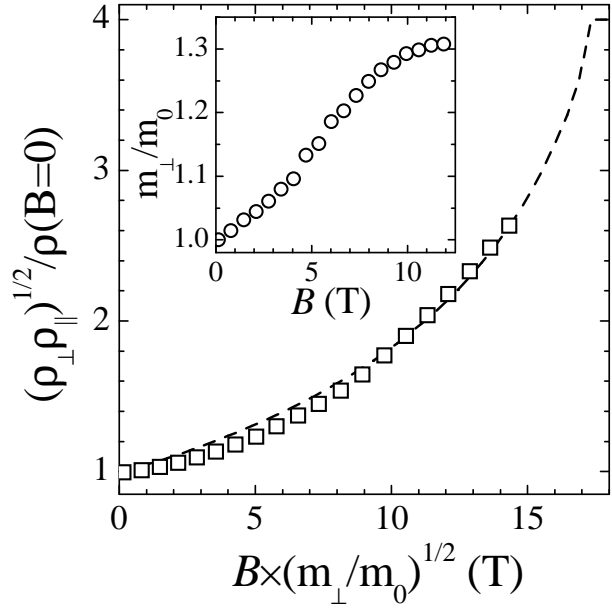}
\end{center}
\caption{In the inset the effective mass growth caused by the
orbital effect~\cite{sarma} is shown, extracted from the
longitudinal MR data of Ref.~\cite{khrapai}
($n_S\approx3\times10^{10}cm^{-2}$) as described in the text. The
spin effect contribution is shown in the figure body. The fit to
theoretical dependence~\cite{dolgold} (dashed line) gives full
spin polarization field in the absence of orbital effect
$B^0_P\approx17.3T$ corresponding to a Land\'e factor
$g^*\approx2.1$ \label{fig}}

\end{figure}

In the inset to Fig.\ref{fig} we show the effective mass growth
extracted in the above manner from the $\rho_\perp,\rho_\parallel$
data of Ref.~\cite{khrapai}.
The effective mass in perpendicular to field direction grows by
about 30\% in moderate fields, as caused by the orbital
effect~\cite{sarma}. For such a slight deformation the utilized
approximation of elliptic Fermi surface should work reasonably, in
contrast to general case~\cite{merkt}. The partial spin
polarization has major effect on MR, as reflected by its small
anisotropy~\cite{khrapai}, and leads to a roughly threefold
resistance increase as is shown in the body of Fig.\ref{fig}.
The full spin polarization has not been reached in~\cite{khrapai}
that's why the saturation~\cite{dolgold} of the geometrical mean of the parallel and perpendicular to inplane field
resistivities predicted by~(\ref{spin}) is not seen in Fig.\ref{fig}.
Fitting to calculation~\cite{dolgold} we obtain the $B^0_P$ value
17.3~T which corresponds to a Land\'e factor $g^*\approx2.1$ at
$n_S\approx3\times10^{11}cm^{-2}$, in agreement with previous studies~\cite{khrapai,shayegan}.

We would like to add a note here, concerning the applicability of the MR data analysis presented above to the
real interacting quasi 2D systems. Apart from the so far considered single particle
effect of Fermi surface deformation, the squeezing of the 2D layer by the parallel magnetic field additionally
changes the form-factors of Coulomb interaction between particles~\cite{afs} and increases the Wigner-Seitz ratio~\cite{tutuc2003},
which can in principle lead to the renormalization of the zero field effective mass and g-factor~\cite{guiliani}.
In presence of such many-body effects the spin effect contribution~\cite{dolgold} cannot be extracted with formula~(\ref{spin}). Experimentally this means that the geometrical mean of resistivities $\sqrt{\rho_\perp(B)\cdot\rho_\parallel(B)}$
does not saturate upon the reach of full spin polarization.
The inplane magnetoresistance anisotropy, however, should still give the anisotropy of
effective mass $\rho_\perp/\rho_\parallel=m_\perp/m_\parallel$, similar to the single-particle picture,
although the independent measurement
is required to find the full spin polarization field $B^0_P$~\cite{zhu,tutuc2003}.

In conclusion, we have derived the linear response function and
dielectric function of 2D Fermi gas with anisotropic effective
mass. In the low density limit the screened charged impurity
potential is shown to possess the same symmetry as the kinetic
energy as a function of wave vector. As a result the elastic
transport relaxation time $\tau$ is isotropic in this limit, even
if the 2D system is partially spin polarized. This finding allows
us to separate for the first time the influence of the
orbital~\cite{sarma} and spin~\cite{dolgold} effects on the
inplane magnetoresistance of a diluted quasi two dimensional
system.

The author would like to thank V.T. Dolgopolov, S.V. Iordanski, A.A. Shashkin and
A.A. Zhukov for useful discussions and acknowledges support from
RFBR, and from the Russian Ministry of Sciences under the
Programmes of ''Nanostructures'' and ''Mesoscopics''.

\begin{thebibliography}{99}
\bibitem{remark2} The low density limit determined by inequality $k_x^F,k_y^F\ll q_{TF}$
is essentially the limit of strong interactions $r_S\gg1$, where
$r_S$ is a Wigner-Seitz radius~\cite{afs}. Strictly speaking, the
utilized random phase approximation is expected to fail in this
limit and the local field corrections should be taken into
account~\cite{pines}. The simplified treatment we use is easily
solved and seems to catch a major effect of Fermi surface
anisotropy.
\bibitem{remark} One can easily see that these conclusions are also valid for the elastic scattering life time, which corresponds to omitting the second term in the integrand of~(\ref{time1}).
\end{thebibliography}
\end{document}